\documentclass[twocolumn,secnumarabic,amssymb, amsmath, nofootinbib,tightenlines,showpacs, aps, prl]{revtex4}

\usepackage{bm}%

\begin{document}
\input{epsf}
\def\Fig#1{Figure \ref{#1}}
\def\Eq#1{Eq.~(\ref{#1})}
\newcommand{\B}[1]{{\bm{#1}}}
\newcommand{\C}[1]{{\mathcal{#1}}}
\newcommand{\pa}{\partial}
\draft
\title{The Breakdown of Linear Elastic Fracture Mechanics near the Tip of a Rapid Crack}
\author{Ariel Livne, Eran Bouchbinder and Jay Fineberg}
\address{Racah Institute of Physics, Hebrew
University of Jerusalem, Jerusalem 91904, Israel}
\begin{abstract}
We present high resolution measurements of the displacement and
strain fields near the tip of a dynamic (Mode I) crack. The
experiments are performed on polyacrylamide gels, brittle elastomers
whose fracture dynamics mirror those of typical brittle amorphous
materials. Over a wide range of propagation velocities
($0.2\!-\!0.8c_s$), we compare linear elastic fracture mechanics
(LEFM) to the measured near-tip fields. We find that, sufficiently near the tip, the measured stress
intensity factor appears to be non-unique, the crack tip significantly
deviates from its predicted parabolic form, and the strains ahead of
the tip are more singular than the $r^{-1/2}$ divergence predicted
by LEFM. These results show how LEFM breaks down as the crack tip is approached.
\end{abstract}
\pacs{46.50.+a, 62.20.Mk, 89.75.Kd} \maketitle

The behavior of a cracked body under applied stress is of extreme
practical and fundamental importance. The accepted approach to
describing crack dynamics is linear elastic fracture mechanics
(LEFM) \cite{Freund.90}. This framework assumes that a material is
linearly elastic with all nonlinear and dissipative processes well
confined to the near tip vicinity. LEFM provides a full description
of the elastic fields surrounding the tip of a single crack, whether
static or propagating. Open questions, however, such as criteria for
crack path selection \cite{99MF}, the origin of dynamic
instabilities (micro-branching) \cite{99MF} and oscillations of a
single crack at high velocities \cite{Oscillations} underline the
need for a more detailed understanding of the near tip behavior in
dynamic fracture. Studies of this region have been limited mainly to
static cracks \cite{Ravi-Chandar book,Bouchaud}, as detailed
measurement of a microscopic region which may move at speeds
approaching sound speeds entails enormous difficulties.

Here we report on recent experimental results in the dynamic
fracture of brittle polyacrylamide gels. In these soft materials,
fracture dynamics are identical to those observed in standard
brittle amorphous materials, but crack velocities, which scale with
material sound speeds, are nearly three orders of magnitude lower
\cite{Livne}. Thus, detailed visualization of the fields in the
near-tip region within an effectively 2D medium becomes possible.
The measurements provide a detailed description of the near-tip fields. Quantitative comparison with LEFM delineates its domain of validity. These findings suggest the importance of nonlinear elasticity as the crack tip is approached.

The brittle gels used in these experiments are transparent,
neo-Hookean, incompressible elastomers that were used in
\cite{Oscillations}. They are composed of 13.8\% total monomer and
2.6\% bis-acrylamide cross-linker concentrations. The shear
($\mu\!=\!35.2\!\pm\!1.4$kPa) and Young's ($E\!=\!3\mu$) moduli of
these gels yield shear and longitudinal wave speeds of
$c_s\!=\!5.90\!\pm\!0.15$m/s and $c_L\!=\!11.8\!\pm\!0.3$m/s. Their
typical dimensions were ($X \!\times\! Y\! \times\! Z$) $130\!
\times\! 125\! \times\! 0.2$mm where $X$, $Y,$ and $Z$ are,
respectively, the propagation, loading, and gel thickness
directions. Reducing the gel thickness suppresses micro-branching
and enables single-crack states to attain high velocities in
effectively 2D media \cite{Oscillations}.

The gels were cast between two flat glass plates. The face of one
plate was randomly scratched with No. 600 Alundum lapping powder.
These scratches, of $16\mu$m mean depth, were imprinted on one of
the gel faces. The resulting scratch pattern was used as a ``tracer
field" for visualization of the displacement field, and did not
affect the crack dynamics.

Experiments were performed as in \cite{Livne} by imposing uniaxial
(Mode I) loading via constant displacement in the vertical ($Y$)
direction. Once the desired stress was reached, a seed crack was
imposed at the sample's edge, midway between the vertical
boundaries. As Fig. 1(a) shows, the displacement field of the
dynamic crack that emerged was visualized at the center of the gel
by a high-speed  (485Hz frame rate) camera that was focused on a
fixed 15.4mm ($X$) by 12.3mm ($Y$) region. The camera's
$1280\!\times\!1024$ pixel resolution enabled a $12\mu$m spatial
resolution. Image blur was limited to less than a single pixel by
using a $2.8\mu$s exposure time.

A fundamental quantity of fracture mechanics is the displacement
field $\B u$, defined by $\B x'\!=\!\B x\!+\!\B u(\B x)$, where $\B
x$ is an un-deformed `rest' configuration and $\B x'$ a deformed
one. We measured this quantity as follows. We first visualized the
scratch pattern by shadowgraphy \cite{shadowgraphy}. We then found
the displacement field between a photograph taken immediately
preceding fracture initiation (reference frame), and photographs
capturing traveling cracks. This was done by means of a particle
tracking technique where the scratches were tracked by
cross-correlating small boxes (10-20 pixels in length, at 5 pixel
intervals) from the reference frame with corresponding regions
deformed by the crack's passage. The maximum correlation for each
box provided a sub-pixel measurement of the displacement field
generated by the crack. Correcting for the uniform stretch of the
reference frame yielded $\B u(\B x)$ in the rest frame, thereby
enabling a direct comparison to LEFM.

\begin{figure}
\centerline{\epsfxsize = 8.5cm \epsffile{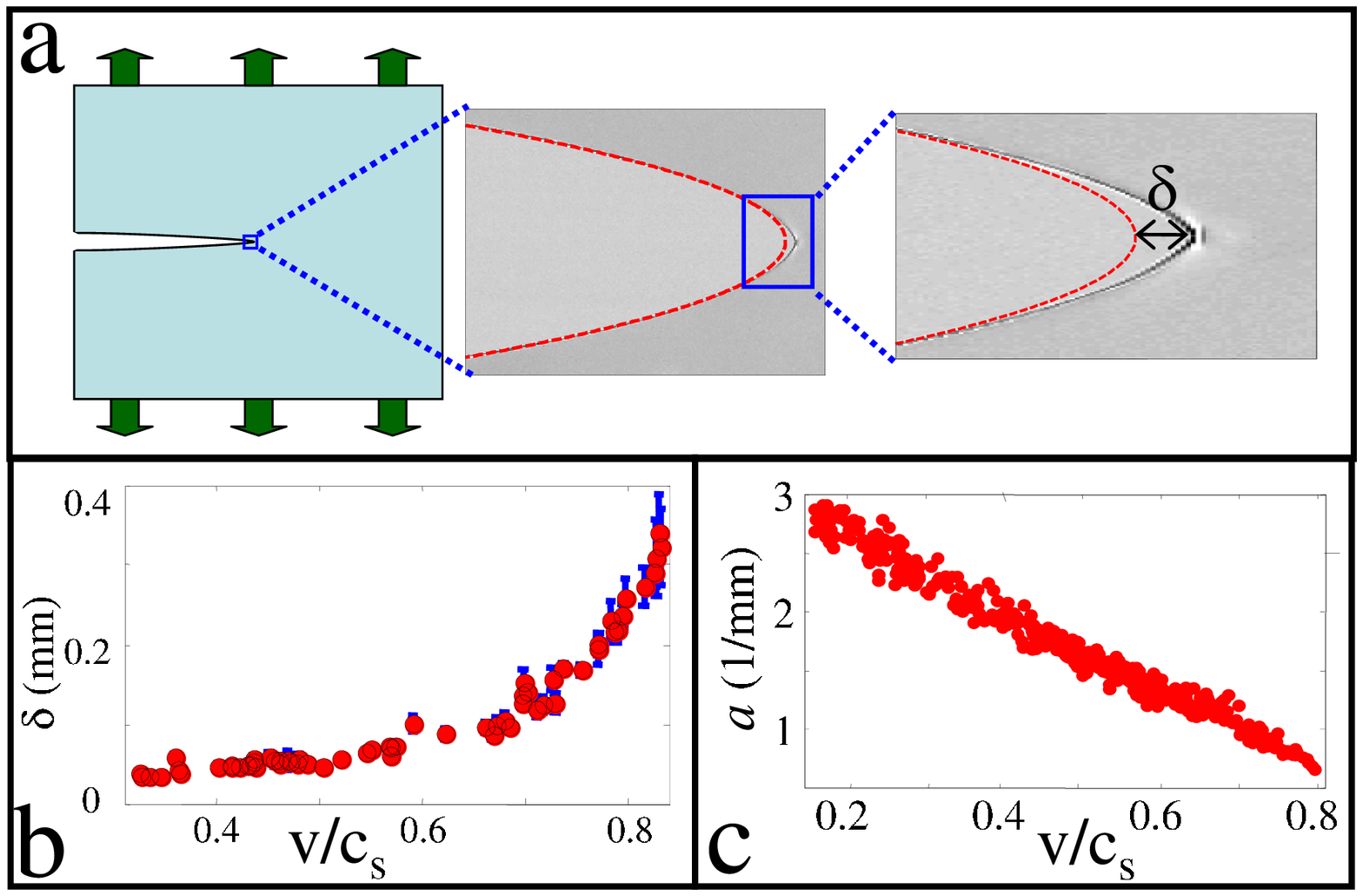}} \caption{(Color
online) (a) (left) A single crack traveling from left to right in a
(to scale) uniaxially loaded sample. (center) Photograph (here
$5.4\!\times\!5.2$mm) of the profile at the tip of a dynamic crack
($v\!\sim\!0.73c_s$). A parabolic fit (dashed red curve) agrees well
with the crack profile for a distance larger than 5mm. (right) A
closeup view ($1.0\! \times\! 1.8 $mm) reveals a deviation,
$\delta$, between  the tip of the parabola and the crack tip. (b)
$\delta$  and (c) the parabola curvature, $a$, as functions of
$v/c_s$.}\label{Ariel3}
\end{figure}

Let us first consider the crack tip opening displacement (CTOD),
which is the clearest manifestation of the displacement field in the
cracked body. LEFM predicts a parabolic CTOD:
\begin{eqnarray}
x'={-\frac{32\pi(1+T/E)}{[3K_I/E{\cdot}\Omega_y(\theta\!=\!\pi,v)]^{2}}}{\cdot}y'^2
\equiv-a{\cdot}y'^2. \label{Eq1}
\end{eqnarray}
We define moving frame coordinates, $(r,\theta)$
($r\!=\!\sqrt{(x-vt)^2+y^2}$ and $\theta\!=\!\tan^{-1}[y/(x-vt)]$),
centered at the crack tip where $\theta\!=\!0$ is the propagation
direction. In Eq.(\ref{Eq1}), $K_I$ is the stress intensity factor,
$T$ is the sub-leading correction known as the ``T-stress'' and
$\Omega_{x,y}(\theta;v)$ are universal functions \cite{following} of
$\theta$ and the crack velocity, $v$. As $T/E \ll 1$, the curvature,
$a$, of the parabola provides a direct measurement of $K_I$. This
important quantity, according to LEFM, wholly dictates the behavior
of a moving crack.

To obtain $K_I$, we extracted the crack opening profiles from the
photographs and fitted them to parabolic forms over 1-2mm from the
crack tip, which is within the range where LEFM might be expected to
be relevant. The parabolic profiles obtained from the fitted data
agreed well with CTOD profiles at significantly larger scales,
indicating their robustness. For example, the fitted parabola in
Fig. 1(a) is indistinguishable from the CTOD profile over 5mm. The
crack-tip curvature is seen to be a decreasing function of the
velocity [Fig. 1(c)]. Interestingly, it is well described by a
linear function with its extrapolated zero intercept at a velocity
close to $c_s$.

A closer look, however, [Fig. 1(a) right] reveals that a noticeable
deviation between the fit and the crack edges occurs in the near
vicinity of the tip. This deviation, $\delta$, increases rapidly
with crack velocity, from $\sim\!30\mu$m at low velocities
($\sim\!0.2c_s$) up to $\sim\!300\mu$m at high velocities
($\sim\!0.8c_s$) [Fig. 1(b)]. The near-tip region is one of
considerably higher curvature than the ``far-field" regions captured
by the parabolic form.  Attempts to force the tip of the parabola to
coincide with the measured crack tip yielded terrible compatibility
with the data at {\em all} scales. We are therefore led to the
conclusion that the near-tip region described by $\delta$ is {\em
not} described by LEFM. We will use $\delta$ to characterize the
scale of this divergence. Although somewhat arbitrary, other
length-scales (e.g. the distance between the crack tip and the point
where the divergence begins) display similar functional behavior.

The near-tip region characterized by $\delta$  can not be explained
in the framework of LEFM. A natural explanation for this region of
deviation would be to associate it with the ``process zone'' scale,
where elasticity breaks down due to extreme stresses, giving way to
plastic deformations and fracture itself. LEFM avoids any treatment
of this zone and regards it as a singular point under the conditions
of small-scale yielding \cite{Freund.90}. Another possible signature
of the process zone may be the white region visible at the tip of
the crack [Fig. 1(a) right]. In this region, the high strains induce
large material deformations giving rise to lensing effects. In
pictures, as this one, where the camera's focal plane is set
slightly below the gel (between the gel and the light source) we
observe increased light intensity in the vicinity of the crack. When
the focal plane is slightly above the gel plane (between the gel and
the camera), the same region becomes black. Hence the near tip
region behaves as a diverging lens.

\begin{figure}
\centerline{\epsfxsize = 8.5cm \epsffile{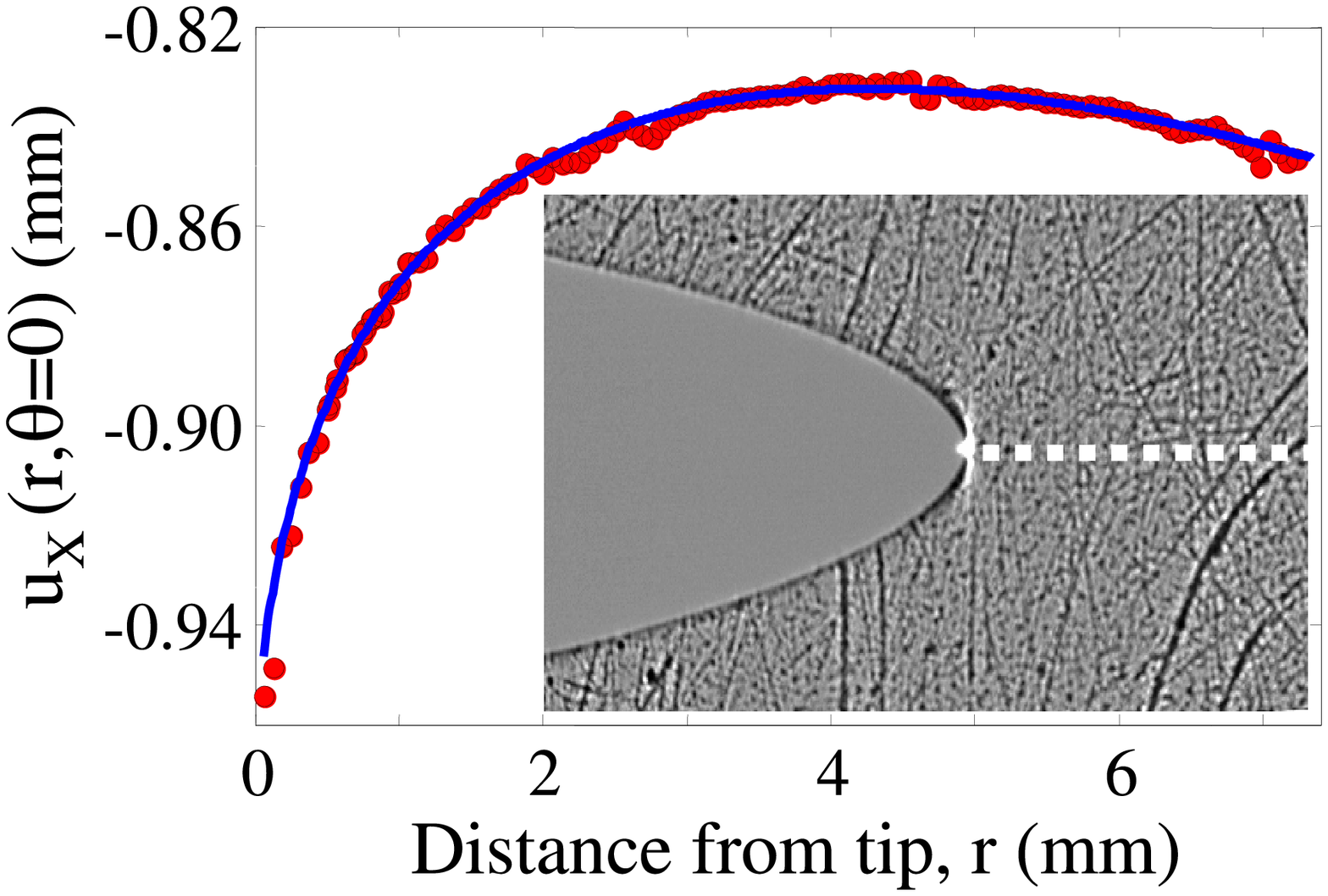}} \caption{(Color
online) Scratch patterns (inset) imprinted on the gels are used to
measure the displacement field surrounding the crack tip via
particle tracking. The measured $u_x$ ($v\!=\!0.37c_s$) along the
$\theta\!=\!0$ axis of symmetry (dashed white line) is fitted to Eq.
(\ref{Eq2}). The values of $u_x$ are accurate up to a small additive
constant ($<\!0.05$mm).}\label{Ariel3}
\end{figure}

Let us now consider the displacement field $\B u(r,\theta)$ around
the crack tip at other $\theta$'s. The best defined direction is
along the axis of propagation directly ahead of the crack
($\theta\!=\!0$). In Fig. 2 we present an example of the measured
displacements along this symmetry axis, obtained using the particle
tracking method explained earlier. We first notice that the
displacements are negative, since particles ahead of the crack are
pulled towards it. Comparing these measurements to the predicted
LEFM functional form \cite{Freund.90}
\begin{equation}
u_x(r,0) =
\frac{3\Omega_x(0,v)}{4\sqrt{2\pi}}\frac{K_I}{E}\cdot\sqrt{r} +
\frac{T}{E}\cdot r + Const, \label{Eq2}
\end{equation}
we obtain excellent agreement at all measured scales, from distances
{\em well beyond} $\delta(v)$ (7mm in Fig. 2) to the closest
near-tip vicinity that we are able to measure \cite{Comment 2 tip
location for u_x fit}. Eq. (\ref{Eq2}) contains the contribution of
the T-stress as well, since the K-field alone is insufficient to
describe the data. Thus in the region of study we have $K$-T
dominance instead of the typically assumed $K$-dominance. The next
order term in the $u_x$ expansion ($r^{3/2}$) is the first to
incorporate dynamic corrections due to non-steady state
contributions like $\dot{v}$ and $\dot{K}_I$ \cite{Freund.90}. We
found no evidence in our data for the existence of such
contributions.

\begin{figure}
\centerline{\epsfxsize = 8.5cm \epsffile{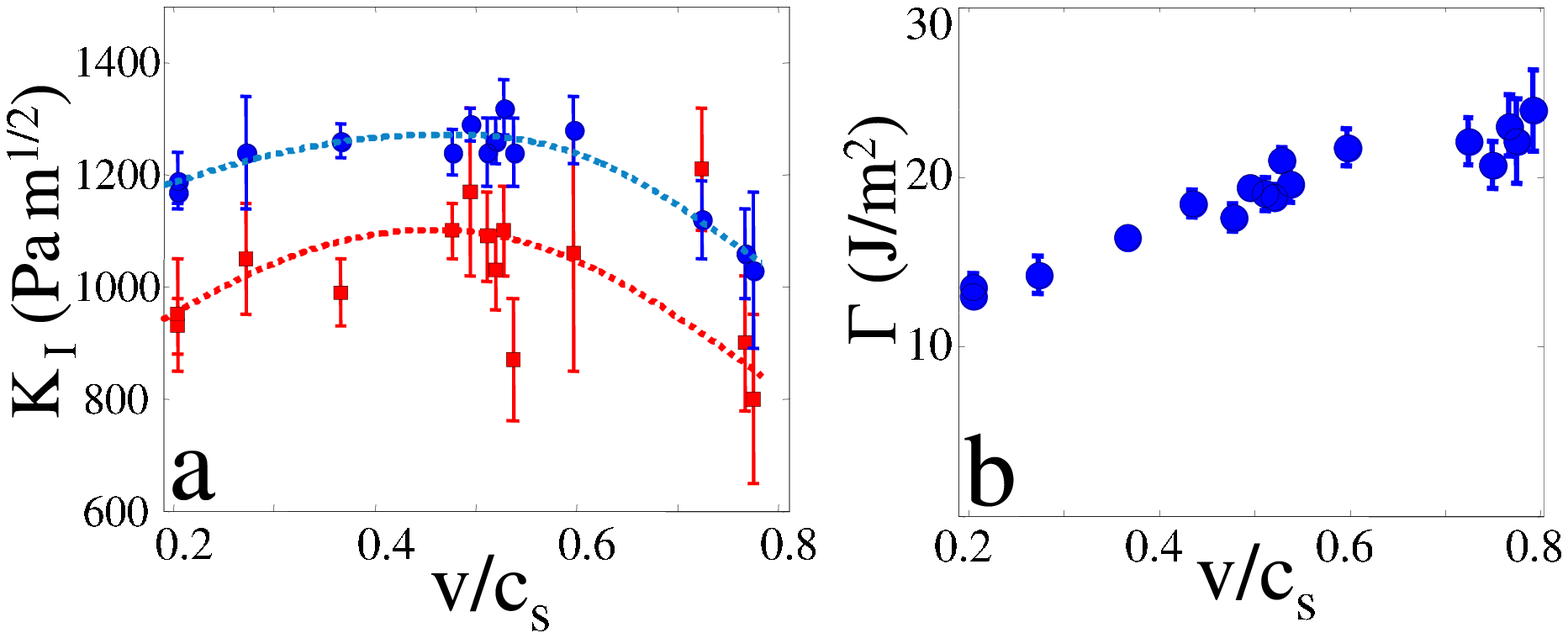}} \caption{(Color
online) (a) The stress intensity factor, $K_I$, is extracted from
crack opening displacement (CTOD) using Eq. (\ref{Eq1}) (circles)
and from $u_x(r,\theta\!=\!0)$ measurements using Eq. (\ref{Eq2})
(squares). The 20$\%$ difference between the two is well beyond
measurement uncertainties. Dashed lines are to guide the eye. (b)
The fracture energy, $\Gamma$, calculated using $K_I$ (derived from
the CTOD using Eqs. (\ref{Eq1}) and (\ref{Eq3})). The weak increase
of $\Gamma$ with $v$ relative to $\delta$ implies that $\delta$ is
due to nonlinear elasticity, rather than dissipative processes.}
\label{Ariel3}
\end{figure}

We are now in a position to compare the values of $K_I$ extracted
from the two different methods:  the displacement field along the
symmetry axis (using Eq. (\ref{Eq2})) and the CTOD measurements
(using Eq. (\ref{Eq1})). We expect to find quantitative agreement
between the two measurements, as LEFM predicts that $K_I$ is unique
and has no angular dependence. Curves of  $K_I(v)$ extracted from
both measurements are presented in Fig. 3(a). Although both describe
a similar functional profile: $K_I$ increases slowly until reaching
a peak at $v\!\sim\!0.5c_s$ before decreasing again,  surprisingly,
the two $K_I(v)$ curves {\em quantitatively} differ. The systematic
20\% difference in the values of $K_I$ is much larger than our
measurement error and does not appear to be $v$-dependent. As our
measured values of $T/E$ are less than a few percent, any
uncertainty in their value could not explain this large deviation in
$K_I$.

Unlike the quantitative discrepancy, the qualitative behavior of
$K_I$ is easily understood. According to LEFM, $K_I$ factorizes into
a universal dynamic component, which depends solely on the velocity
and a geometric component, which depends on the loading and
dimensions of the stressed object. While the first is a decreasing
function of $v$, the latter increases with $v$ through the loading.
The non-monotonicity of $K_I(v)$ is due to the competition between
these two components.

Using $K_I(v)$, we can now compute the material's fracture energy,
$\Gamma$ \cite{Freund.90}
 \begin{equation}
 \Gamma(v) = K_I^2(v)A_I(v)/E
\label{Eq3}
\end{equation}
where $A_I(v)$ is a known \cite{Freund.90} function of $v$. Using
the values of $K_I(v)$ obtained from the CTOD, we observe [Fig.
3(b)] that $\Gamma$ is a slowly monotonically increasing function of
$v$ whose value is approximately 20J/m$^2$. Similar values have been
reported for other gels \cite{Fracture Energy of gels}. $\Gamma$ is
a measure of the dissipation involved in crack propagation. It
cannot, in general, be estimated from first principles, but is a
material property which must be measured.

We are now in a position to understand the nature of the processes
that govern the deviations $\delta$ from the parabolic CTOD. In
principle, $\delta$ can either be the scale of dissipative processes
(i.e. the process zone) or it may be associated with {\em nonlinear}
elastic processes. If the former were correct, one would expect
$\Gamma(v)$ and $\delta(v)$ to have a similar functional dependence.
Comparison over the velocity range of $0.4\!-\!0.8c_s$ reveals only
a modest $\sim\!30 \%$ increase in $\Gamma$, while $\delta$
increases by over $\sim\!400\%$. This large contrast indicates that
nonlinear (non-dissipative) elastic processes are dominant in the
near-tip region at scales that are significantly larger than
dissipative ones.

\begin{figure}
\centerline{\epsfxsize = 8.5cm \epsffile{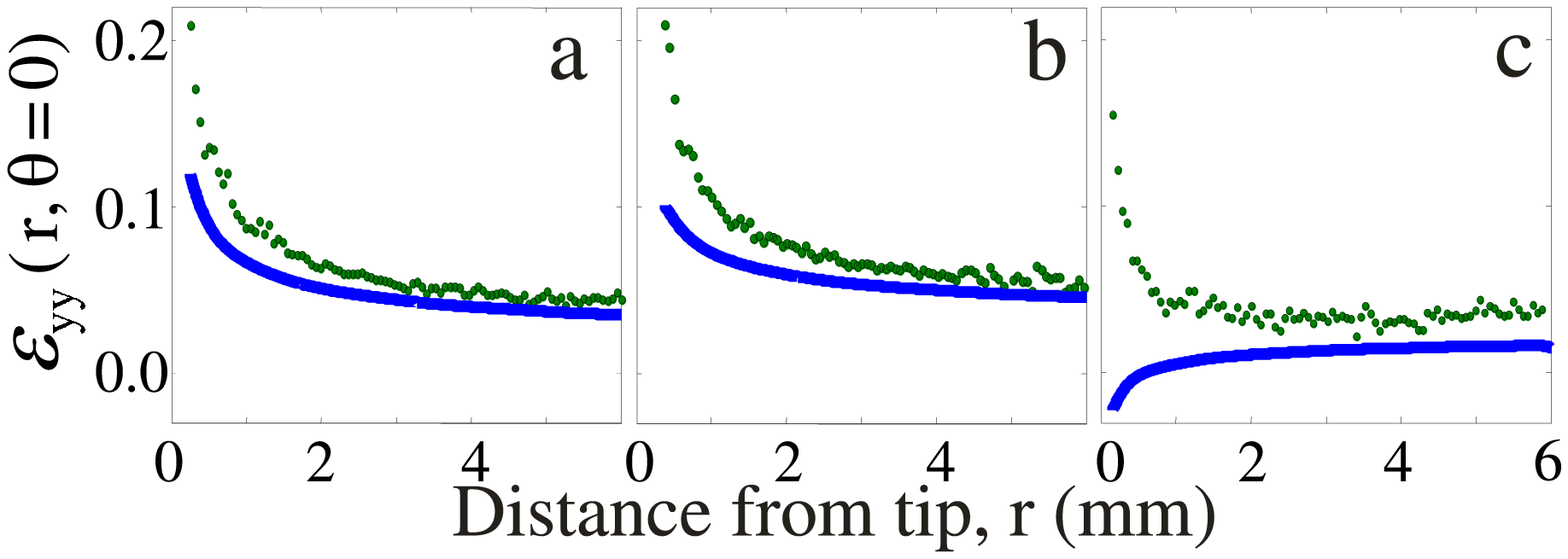}} \caption{(Color
online) The measured strain, $\varepsilon_{yy}(r,\theta\!=\!0)$,
(dots) is compared to the theoretical (LEFM) prediction (solid
curve): $[g_y(\theta\!=\!0,v)\!\cdot\!K_I/E]/\sqrt{r}-0.5T/E$ where
$K_I$ and $T$ are taken from the $u_x$ fit (cf. Fig. 2) and $g_y$ is
a known universal function \cite{Freund.90}. The discrepancy between
the two increases with the crack velocity: (a) $v\!=\!0.20c_s$, (b)
$v\!=\!0.53c_s$, (c) $v\!=\!0.78c_s$. For the higher velocity (c),
LEFM predicts a {\em negative} strain (compression) ahead of the
crack tip.}\label{Ariel3}
\end{figure}

The large discrepancy between the two values of $K_I$ derived from
different components of the {\em same} displacement field, motivated
us to examine also the deformation in the y-direction at
$\theta\!=\!0$. Since for Mode I symmetry, $u_y(r,0)\!=\!0$, we
instead use the strain $\varepsilon_{yy}(r,\theta=0)\!=\!\partial_y
u_y(r,0)$. In Fig. 4 we compare $\varepsilon_{yy}(r,0)$ at three
different velocities, from $20\%$ to $\sim\!80\%$ of $c_s$, to the
LEFM predictions \cite{Comment 4 how the strain was evaluated}. At
all velocities we measure strains of $\sim\!0.2$-$0.3$, before
encountering measurement limitations \cite{Comment 2 tip location
for u_x fit} when too close ($\sim\! 200\mu$m) to the tip. LEFM
predictions for $\varepsilon_{yy}(r,0)$ were calculated using the
$K_I$ and T stress values obtained by fitting the $u_x(r,0)$
displacement components at the same velocities, cf. Fig. 2.

In stark contrast to the excellent fits obtained for $u_x$, the
predicted values of $\varepsilon_{yy}$ deviate significantly from
the measured ones. At the lower velocities, the measured strain
increases notably more rapidly than the $r^{-1/2}$ dependence
predicted by LEFM [Fig. 4(a-b)], and attempts to fit the strains
using LEFM fail. At higher velocities [Fig. 4(c)] the deviations
between measured and expected values are even more dramatic.

For $v\!\!>\!\!0.73c_s$ (for incompressible materials) LEFM
predicts a monotonically {\em decreasing} strain which reaches
negative values (compressive strains) as one approaches the crack
tip. On the other hand, our measurements show the strain to be a
positive, monotonically {\em increasing} function that reaches
near-tip values that are similar to those measured at lower
velocities. The measurements make more intuitive sense than the LEFM predictions - tensile fracture occurs under
extension.

The experiments described here present high resolution measurements
of the displacement field surrounding dynamic cracks. We show that
the canonic theory of fracture, LEFM, fails to provide a consistent
description of the experimental data, apparently as a result of
elastic nonlinearities \cite{Buehler03_06}. This, by itself, is
neither surprising nor contradicts LEFM, as these effects may be
conceptually incorporated into one of its key assumptions, the
small-scale yielding condition \cite{Freund.90}.

On the other hand, as fracture occurs precisely {\em at} the
smallest scales, a description of this near-tip region is important.
Above, we presented detailed measurements within the nonlinear
region in this type of material. In addition, we demonstrated that
at high $v$, LEFM {\em qualitatively} fails  to describe the tensile extensions measured ahead of the crack tip
cf. Fig. 4(c). This underlines the necessity for a more complete
theoretical understanding of essential nonlinear effects.

Are these observations unique to elastomers or are they more
generally valid? In the accompanying Letter \cite{following} we
develop a weakly nonlinear theory of the dynamic fracture of a
single crack that resolves all of the discrepancies discussed
above. Such nonlinearities are as universal as linear elasticity
and {\em must} be experienced by {\em any} material undergoing
fracture. Therefore, we expect these results to be generally
applicable to any brittle material.

The key features of brittle fracture in the gels considered here are
{\em identical} to those observed in other brittle amorphous
materials, like glassy polymers or structural glasses \cite{Livne}.
This work, however, indicates that LEFM is unable to describe these
features in gels, since the near-tip fields are nonlinearly elastic
at scales encompassing the origin of these effects. Moreover, the
near-tip dissipative processes in gels, brittle plastics and glasses
are as different as their micro-structure. This leads us to conclude
that the nonlinear elastic region that bridges the gap between LEFM
and the process zone must play a critical role in governing the
fracture process. Thus, understanding the dynamics within this
region may be the key to unlocking a plethora of open questions that
are related to the breakdown of single straight cracks. These
include questions of stability \cite{99MF} (e.g. micro-branching and
crack oscillations), crack path selection \cite{99MF}, and 3D
nonlinear focussing (e.g. crack front inertia, front wave nonlinear
structure, and the formation of directed chains of micro-branches
\cite{Livne}).

{\bf Acknowledgements} This research was supported by grant 57/07 of
the Israel Science Foundation. E. B. acknowledges support from the
Horowitz Center for Complexity Science and the Lady Davis Trust.


\begin{thebibliography}{10}
\bibitem{shadowgraphy}
Collimated light (LED source) oriented normal to the gel faces was
deflected by the scratches imprinted on the gel surface. Optimal
contrast was obtained by setting the focal plane of the camera
slightly below that of the gel.
\bibitem{Comment 2 tip location for u_x fit}
The measured tip location is used in the fitted function. Note that
we are unable to measure displacements at points closer than
$\sim\!200\mu$m from the tip, due to the extreme strains and lensing
effects that dominate there.
\bibitem{Comment 4 how the strain was evaluated}
We evaluated $\varepsilon_{yy}(r,\theta\!=\!0)$ in the following
manner: for every distance $r$ from the crack tip along
$\theta\!=\!0$, $\varepsilon_{yy}(r,\theta\!=\!0)$ was obtained from
the linear term of a 3rd order polynomial fit of $u_y$ along $y$
(without the quadratic term), where $\leq$16 measured values of
$u_y$ above and below the symmetry axis were used.
\end{thebibliography}
\end{document}